## The role of Hume-Rothery's rules play in the MAX phases formability

Yiming Zhang[1], Zeyu Mao[1, 2], Qi Han[3], Youbing Li[1], Mian Li[1], Shiyu Du[1*], Zhifang Chai[1], Qing Huang[1**]

1. *Engineering Laboratory of Advanced Energy Materials, Ningbo Institute of Materials Technology & Engineering, Chinese Academy of Sciences, Ningbo, China*
2. *School of Materials Science and Engineering (MSE), Xi'an Jiaotong University*
3. *College of Science and Technology, Ningbo University, Ningbo, China*

## Abstract

MAX phases are a family of layered, hexagonal-structure ternary carbides or nitrides of a transitional metal and an A-group element. What makes this type of material fascinating and potentially useful is their remarkable combinations of metallic and ceramic characteristics; as well as the indispensable role in 'top-down' synthesis of their 2D counterparts, MXenes. To enhance the efficiency in the successful search for potential novel MAX phases, the main efforts could go toward creating an information-prediction system incorporating all MAX phases' databases, as well as generally valid principles and the high-quality regularities. In this work, we employ structure mapping methodology, which has shown its merit of being useful guides in materials design, with Hume-Rothery parameters to provide guiding principles in the search of novel MAX phases. The formable/non-formable data on MAX phases can be ordered within a two-dimensional plot by using proposed expression of geometrical and electron concentration factors.

[*] Corresponding author. Tel/Fax: +86-(0)574-87602759, e-mail address: dushiyu@nimte.ac.cn

[**] Corresponding author. Tel/Fax: +86-(0)574-87602759, e-mail address: huangqing@nimte.ac.cn

Postal address: Ningbo Institute of Materials Technology & Engineering, Chinese Academy of Sciences. No. 1219 Zhongguan West Road, Zhenhai District, Ningbo, Zhejiang Province, P.R. China, 315201

## 1. Introduction

The $M_{n+1}AX_n$ (or MAX phases), as the most common precursors for MXenes, are layered machinable ternary carbides and nitrides; where "M" is an early transition metal, "A" is an A group element (mostly from groups 13 and 14), "X" is C and/or N, and n = 1 to 3 [1-3]. During the past, this type of material have attracted considerable attention due to **1)** their combination of both metallic and ceramic properties, **2)** the ease by which one can modify their chemistry, while keeping the structures same, and **3)** the discovery of both in- and out-of-plane ordered phases that opens the door to the discovery of many more [4-6]. Currently, 14 *M*-elements and 20 *A*-elements have been introduced into various end-member MAX phases, resulting more than 100 MAX compositions synthesized to date (shown in Figure 1). In order to explore novel MAX phases efficiently, *ab initio* calculations have been carried out either through calculating whether a compound is stable on an absolute scale, or showing that a given MAX phase is more stable than all other competing phases [3, 7-12]. The predictive power of these tools is evident; yet without time-consuming full DFT calculations, these analyses incapable to reveal any correlations or systematic behaviour [3, 7]. As a result, alternative strategies could be introduced in to develop general chemical design regularities with the hope to facilitate efficient exploration of a broad MAX phases design space.

Structure mapping, a strategy for building classification schemes of homologous compounds from known structures database, has served as a guiding tool for finding stable phases in a bivariate way [13-15]. Through the choice of appropriate coordinates (*i.e.*, the factors governing the stable crystal structures) based on the understandings of physical meaningful correlations between the empirical parameters and the formation phases, one can map clustering of the crystal structure-related data. As mentioned by Pettifor [13], this method has proven important as an initial guide for the search of new compounds with a required structure type; and which compound might take a given structure type could be suggested. There is a long and distinguished tradition of such maps, several examples including the figures developed by Hume-Rothery [16], Mooser-Pearson plots [17], Philips and van Vechten diagrams [18, 19], Zunger maps [20], Villars maps [21-23], Pettifor maps [24, 25] and others [26-33]. In this work, we aim to construct one single 2D structure map to identify the stability field that the $M_{n+1}AX_n$ phases could form; so that all of the MAX phases structure data could be ordered and eventually understood within a microscopic theory. The chemical scales are proposed to provide guiding principles in the search for new MAX phases.

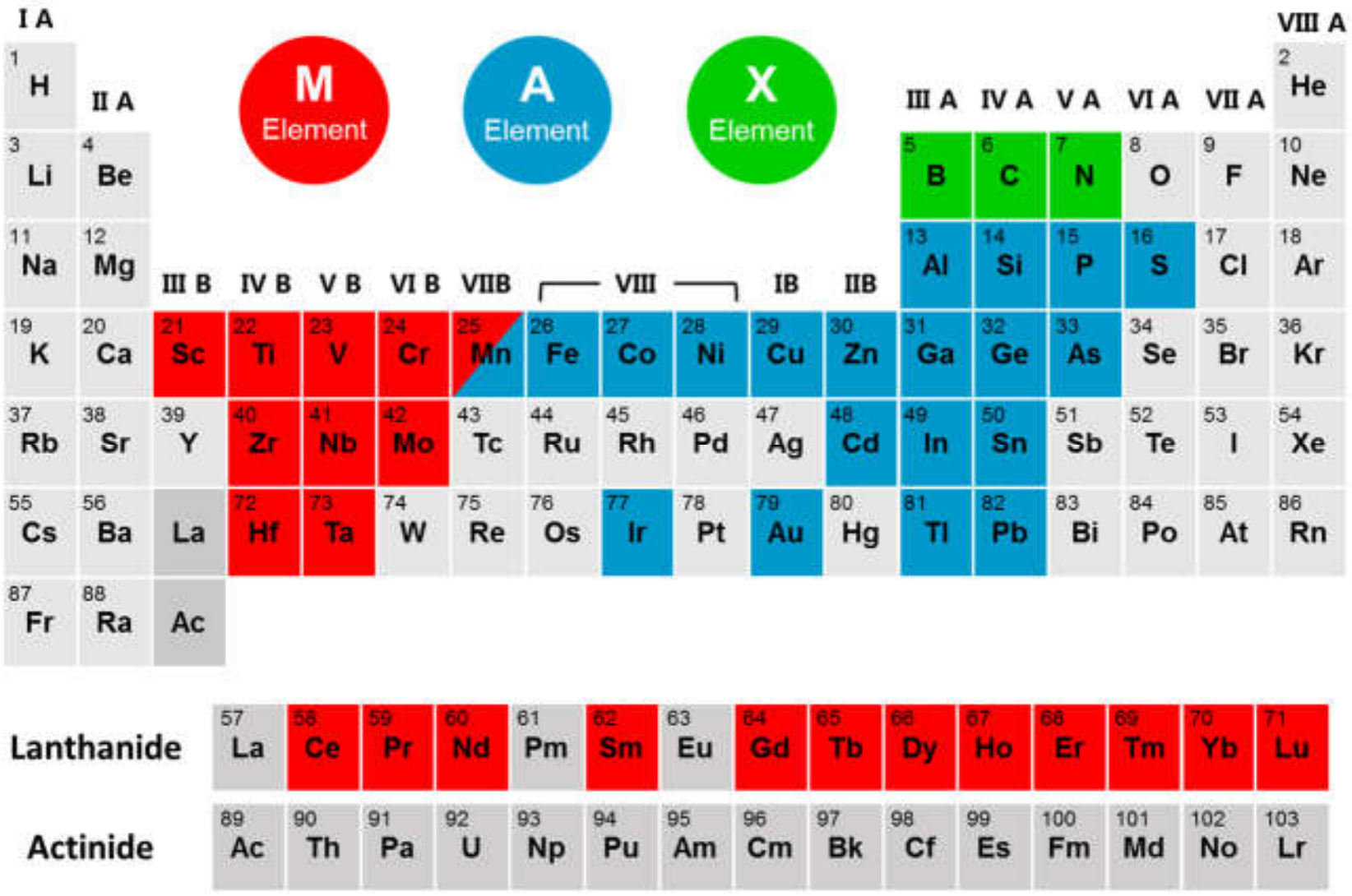

Figure 1 Elements introduced in various ternary MAX phases

## 2. Exploration of the factors regulating MAX phases

It is known that the compound formation, spatial arrangement of atoms, and bond constitution are determined by different groups of factors, namely the 1) geometrical factor; 2) electron concentration factor, 3) electrochemical factor, and 4) angular character of the *s*, *p*, and *d* valence orbitals [13, 34, 35]. The majority of structure mapping constructions are based on taking coordinates that reflect the most important physical factors determining the structural stability of the particular class of compounds under consideration [17, 19-23, 26]. In principle, the classifications of the structures of homologous compounds have to take all of these factors into account; however, it makes the perfectly ordering within a single two-dimensional map challenge with the use of unique physical coordinates. This situation becomes more serious when faced with MAX phases, where the mixture of bonding characteristics complicates the cases. Considering the fact of these factors are not wholly independent with each other, and it is their interplay that determines the structural stability of compound phases. In this work, our goal is to borrow ideas from Hume-Rothery's rules [36-39], which distinguished the factors that control alloying behaviour and influence compound formation over a span of some 100 years, to identify two most important factors for constructing a 2D structure map; in order to capture the dominant factors that control the formability of MAX phases, and then quantify key design variables for their experimental development.

### 2.1 *Determining the types of factors*

The electron concentration is the most prominent of Hume-Rothery parameters in alloy design, which possess intricate ways of electronic interactions among the constituent elements [40]. Referred in his classical monograph [3], Barsoum deemed the formability of MAX phases could be looked from the viewpoint of the average number of valence electrons, $n_{val}$, for a given MAX phase. Taking this inspiration into consideration, the electron concentration factor is adopted as one coordinate in this work. However, the values for this factor differ from the ones mentioned by Barsoum; which are discussed in section 2.2.

It is also known that the unit cell of MAX phase structures is composed of edge-sharing $M_6X$ octahedra with A-site elements located at the centre of trigonal prisms; and the distortion of ideal trigonal prism induced by newly introduced A-site elements would strongly affect the viability of these novel MAX phases. A parameter $p_r$, given by Hug *et al.* [41], has been used to measure this type of distortion; which is determined by inter-atomic distances between M-site and A-site elements, as well as between M-site elements. Based on above analysis, the size factor is chosen as the other coordinate. The expression and values of this factor are discussed within section 2.3.

### 2.2 *Determining the values of electron concentration factor*

There are two different values for defining the electron concentration of individual constituent elements [40]: one is an average number of itinerant electrons per atom (referred to ***e/a***); while the other is the average number of total electrons, including *d*-electrons accommodated in the valence band, which is valence electrons (referred to ***VEC***). Both ***e/a*** and ***VEC*** play crucial roles for the chemical bonding of a solid; and the main difference is that the ***e/a*** is introduced as a concept defined mainly for metallic bonding, whereas the ***VEC*** is locally defined for strong chemical bonding [42, 43].

In MAX phases, it is known that for M-X bonding, the *p-d* interaction between M-site and X-site atoms are quite strong; and the hybridization between the M-site *d* orbitals and the X-site 2*p* orbitals lead to strong covalent bonds. While for M-A bonding, the interactions between the *d* electrons of the M atoms and *p* electrons of the A atoms are weaker than those between the M and X atoms; and A-site *d* electrons do not appear to play a role in bonding [3].

Following above analysis, here we choose ***VEC*** for M-site and X-site elements; and choose ***e/a*** for A-site elements for calculating the electron concentration of MAX phases. The electron concentration of MAX phases can be calculated in Eqn. 1 as follow:

$$Electron\ concentration = \frac{(VEC)_M * n_M + \left(\frac{e}{a}\right)_A * n_A + (VEC)_X * n_X}{n_M + n_A + n_X} \quad \text{Eqn. 1}$$

Where $(VEC)_M$ represents valence electron values of M-site elements; $\left(\frac{e}{a}\right)_A$ represents itinerant electron values of A-site elements; $(VEC)_X$ represents valence electron values of X-site elements; and $n_M$, $n_A$, $n_X$ represent atomic coefficients of M-, A- and X- site elements respectively. The values of valence electrons are collected from general inorganic chemistry textbook [44]; and the itinerant electrons are taken from Mizutani and Sato [42]. Table 1 listed the valence/itinerant electron values adopted for the elements used in this work.

Table 1 The valence/itinerant electron values adopted in this work [42, 44]

| **Elements** | **Valence Electrons** | **Elements** | **Itinerant Electrons** | **Elements** | **Valence Electrons** |
|---|---|---|---|---|---|
| **M** | | **A** | | **X** | |
| Sc | 3 | Al | 3.01 | C | 4 |
| Ti | 4 | Si | 4.00 | N | 5 |
| V | 5 | P | 4.97 | | |
| Cr | 6 | S** | 6.00 | | |
| Mn | 7 | Mn | 1.05 | | |
| Zr | 4 | Fe | 1.05 | | |
| Nb | 5 | Co | 1.03 | | |
| Mo | 6 | Ni | 1.16 | | |
| Lu* | 3 | Cu | 1.00 | | |
| Hf | 4 | Zn | 2.04 | | |
| Ta | 5 | Ga | 3.00 | | |
| | | Ge | 4.05 | | |
| | | As | 4.92 | | |
| | | Pd | 0.96 | | |
| | | Ag | 1.01 | | |
| | | Cd | 2.03 | | |
| | | In | 3.03 | | |
| | | Sn | 3.97 | | |
| | | Ir | 1.60 | | |
| | | Pt | 1.63 | | |
| | | Au | 1.00 | | |
| | | Tl | 3.03 | | |
| | | Pb | 4.00 | | |

* The value of 3 is used for Lu by considering follow the periodic trend

** The work of Mizutani and Sato [42] does not record itinerant electrons for sulphur, here the value of 6 is used for sulphur.

### 2.3 *Determining the values of size factor*

As mentioned in section 2.1, the distortion of ideal trigonal prism induced by A-site elements could be measured by the parameter $p_r$, which is determined by inter-atomic distances between M-site and A-site elements, as well as between M-site elements. In other words, the information about distortion could be described in terms of M/A sizes to a large extent. In Hume-Rothery's rules, the atomic difference ratio of solvent and solute is adopted as size factor to characterize the geometrical mismatch between solvent and solute atoms. Borrowing the idea for the expression of size factor from this rule, the difference between the atomic radii of M-site elements and A-site elements divided by the radii of M-site elements is used as size factor in this work. The values can be calculated in Eqn. 2 as follow:

$$\text{Atomic difference ratio} = \frac{|R_M - R_A|}{R_M} \qquad \text{Eqn. 2}$$

Where $R_M$ and $R_A$ represent atomic radii of M-site elements and A-site elements respectively. The values of atomic radii are taken from CRC Handbook [45]; and Table 2 listed the atomic radius values adopted for the elements used in this work.

Table 2 The atomic radius values adopted in this work (Unit: Å) [45]

| **Elements** | **Atomic Radius** | **Elements** | **Atomic Radius** | **Elements** | **Atomic Radius** |
|---|---|---|---|---|---|
| **M** | | **A** | | **X** | |
| Sc | 1.59 | Al | 1.24 | C | 0.75 |
| Ti | 1.48 | Si | 1.14 | N | 0.71 |
| V | 1.44 | P | 1.09 | | |
| Cr | 1.30 | S | 1.04 | | |
| Mn | 1.29 | Mn | 1.29 | | |
| Zr | 1.64 | Fe | 1.24 | | |
| Nb | 1.56 | Co | 1.18 | | |
| Mo | 1.46 | Ni | 1.17 | | |
| Lu | 1.74 | Cu | 1.22 | | |
| Hf | 1.64 | Zn | 1.20 | | |
| Ta | 1.58 | Ga | 1.23 | | |
| | | Ge | 1.20 | | |
| | | As | 1.20 | | |

| | | |
|---|---|---|
| | Pd | 1.30 |
| | Ag | 1.36 |
| | Cd | 1.40 |
| | In | 1.42 |
| | Sn | 1.40 |
| | Ir | 1.32 |
| | Pt | 1.30 |
| | Au | 1.30 |
| | Tl | 1.44 |
| | Pb | 1.45 |

## 3. Results

In order to construct structure maps, the data of formable and non-formable MAX phases have to be collected. In this work, the MAX phases that have been experimentally synthesized, rather than the viable ones screened by *ab initio* calculations, are collected from Sokol *et al.*[6] as formable phases; and the non-formable ones are recorded from Aryal *et al.* [10], which are screened based on elastic and thermodynamic stability by using *ab initio* calculations. Here it worth noting that there are other works that studied the stabilities of MAX phases systematically, via showing a given phase is more stable than all other competing phases [46, 47]. This computing strategy brings a stricter criterion to screen possible MAX phases; and when done properly, could select MAX phases out showing excellent agreement with experimental observed ones [3]. However, there are several outliers exist within the proposed unformable ones (including $Sc_2InC$, $Nb_2SC$, $Hf_2SnN$, $Hf_2SC$, $Cr_2GeC$, and $Zr_2AlC$, all of which have been synthesized successfully), which would obstruct the distinct classification when introduced in. Further, novel preparation strategies have been developed within authors' group to synthesis MAX phases successfully via avoiding the reactions to form competitive phases [48-50]; and this is the reason why we treat formability, rather than the stability in this work. In contrast, the two criteria employed by Aryal *et al.* [10] are necessary but not sufficient conditions for MAX phase formability; and the non-formable ones screened out by following these two criteria are the ones that would scarcely form. Here, we should reiterate that since the main goal of this study is to capture the principal factors that govern the formability of existing and potential MAX phases, we have to purposely employ the dataset with more confident judgements.

### 3.1 *The structure map construction from traditional 211-MAX phases*

We start our analysis for traditional MAX phases at first. The traditional MAX phases, whose A-site elements mainly from groups 13 and 14, are summarized by Barsoum in his 2013 monograph [3]; and several follow-up research works engaging comprehensive assessment of possible MAX phases focused their study within the scope of these elements in a combinatorial way [10, 46]. Due to most of the traditional MAX phases are 211 phases, the structure map is constructed from them through the calculated values of *electron concentration factor* and *size factor* from Eqn.1 and Eqn. 2 (shown in Figure 2). It is found that there exist two separable fields to which the formable/non-formable MAX phases are belonging. From Figure 2, it can be seen that the chosen factors and their expressions ("*electron concentration factor*" and "*size factor*") in this work has achieved a successful formability/non-formability separation of 211-MAX phases, only with a few exceptions; and the details will be discussed later.

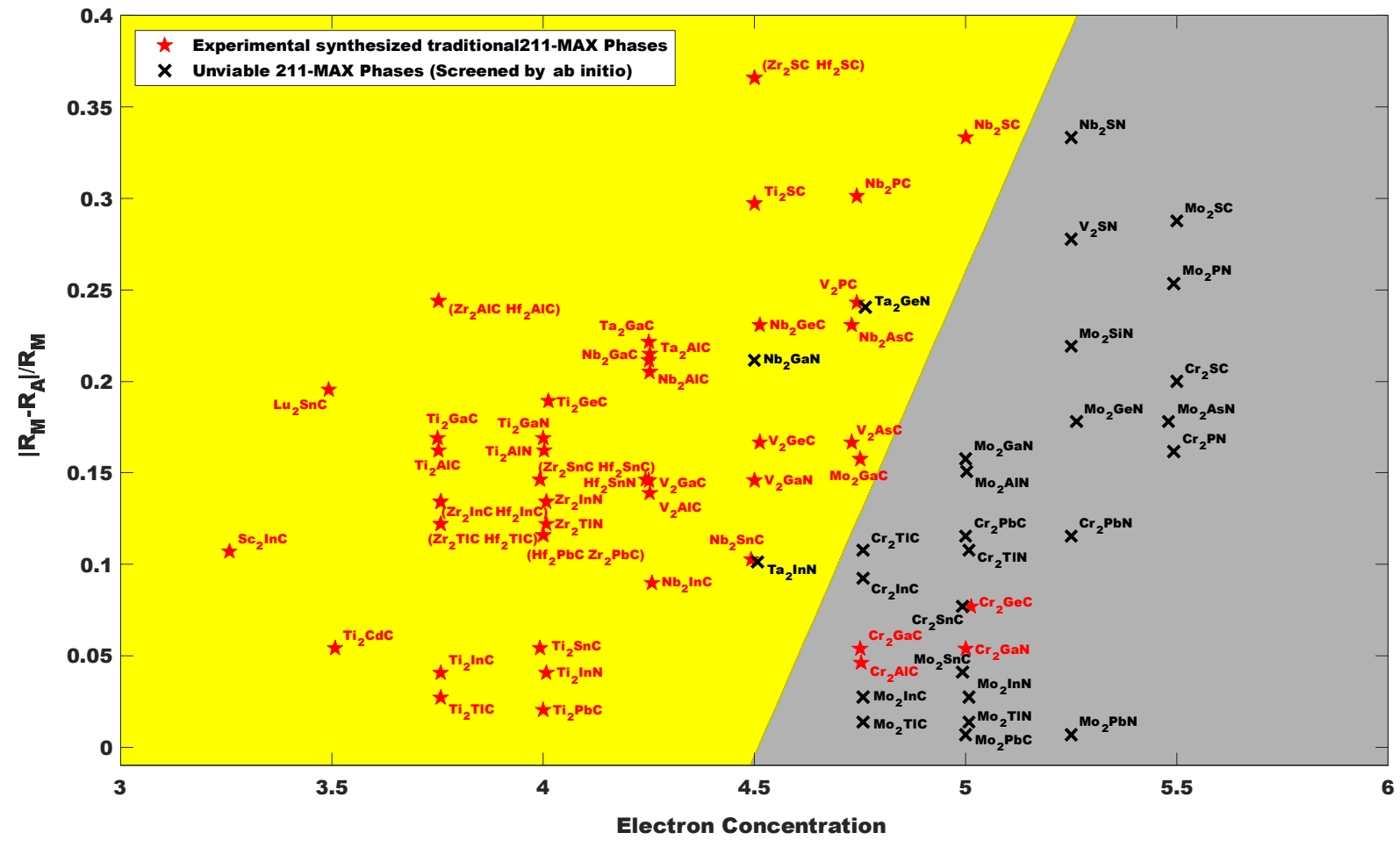

Figure 2 The structure map constructed from traditional 211-MAX phases

### 3.2 *Apply the constructed structure map to traditional 312- and 413- MAX phases*

In the next trail, the constructed structure map is further applied to traditional 312-MAX phases and 413-MAX phases; for examining whether the constructed structure map is the one that could unify the formability/non-formability separation of all traditional MAX phases (shown in Figure 3 (a) and (b) respectively). The results show that the same structure map has capability to locate formable/non-formable 312-MAX and 413-MAX phases into separated fields; which means one single structure map has the capability to classify the formability/non-formability for the whole family of traditional $M_{n+1}AX_n$ (n = 1 to 3) phases. In other words, the formable domain for whole family of traditional MAX phases are framed by the boundaries defined from electron concentration and M/A size difference.

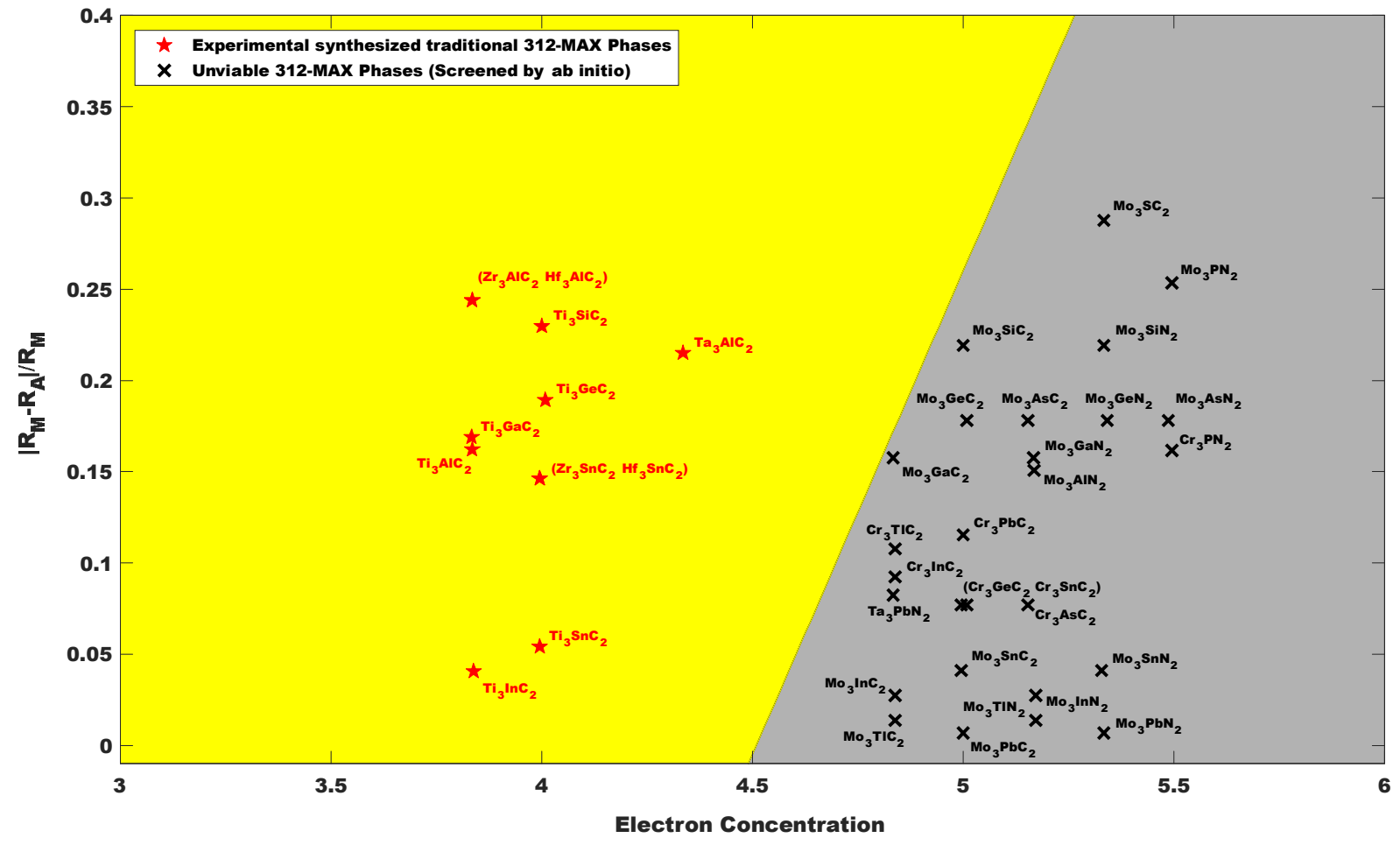

(a)

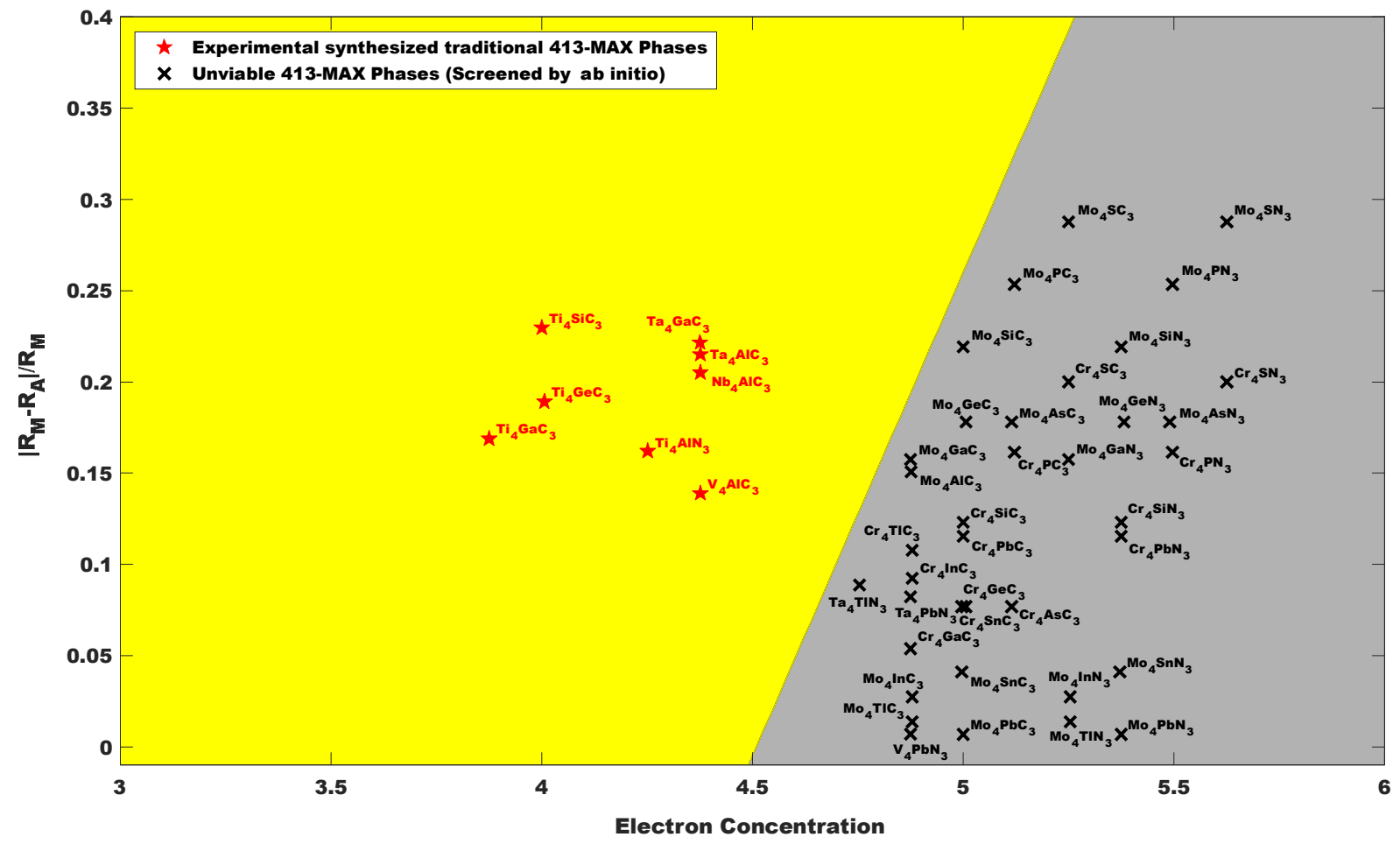

(b)

Figure 3 The constructed structure map applied to (a) traditional 312-MAX phases; (b) traditional 413-MAX phases

## 3.3 *Test the constructed structure map by newly reported MAX phases*

Currently, M/A-site elements replacement in traditional MAX phases by later transition-metals is becoming a buzzing field [48-55]; which opens a door to explore new types of MAX phases that beyond the range of elements mentioned by Barsoum [3]. The quest for generality raises the requirement of validating the developed structure map by using these newly reported novel MAX phases, where the A-site elements have been extended to include Zn, Cu, Au, Ga; and Mn is introduced in as the M-site element. The validation results are shown in Figure 4 (a) and (b). It is found that all the newly reported novel MAX phases, except $Mn_2GaC$, are located within the stability field; which show the generality of the constructed structure map that can be extended to a broad family of $M_{n+1}AX_n$ phases.

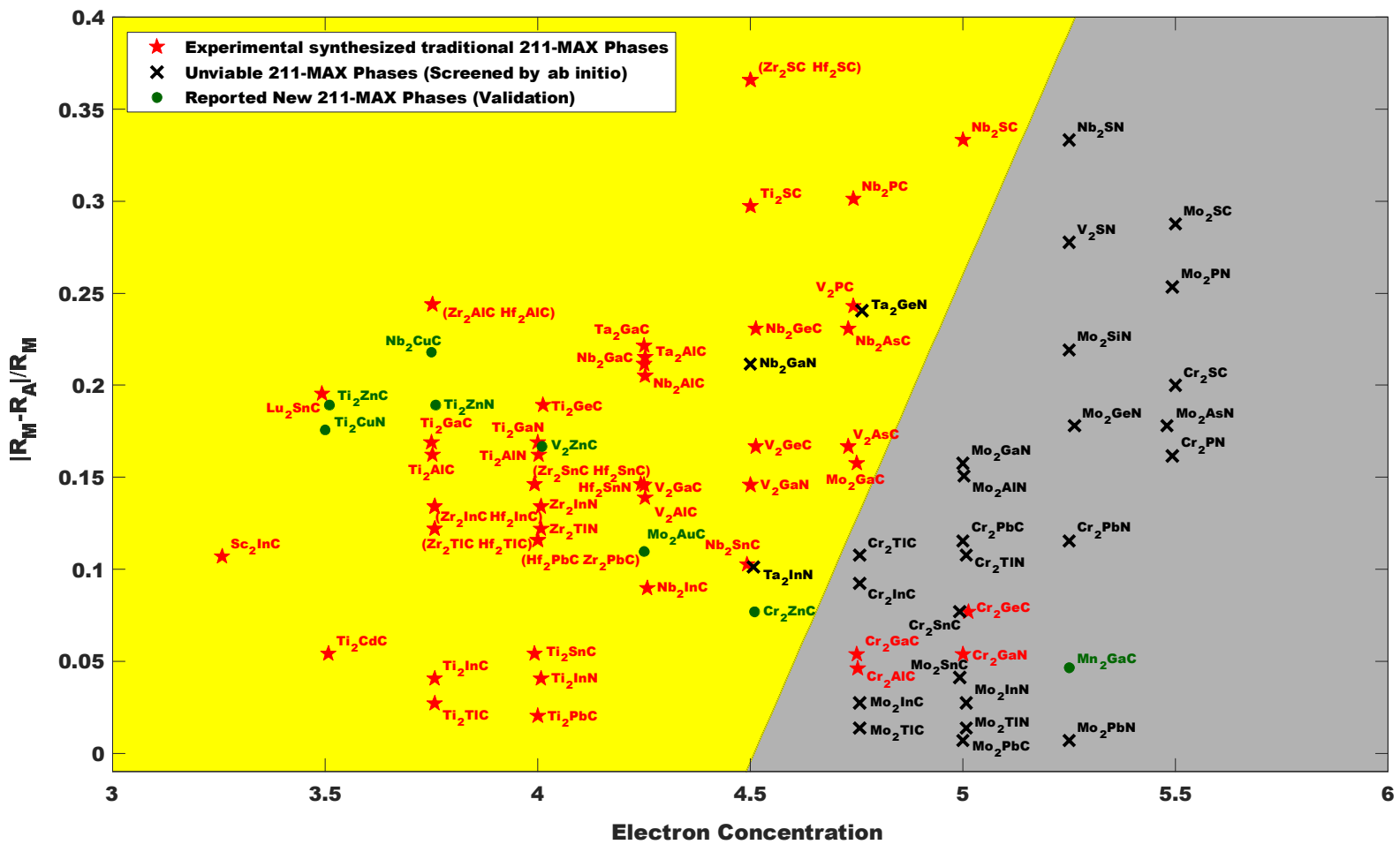

(a)

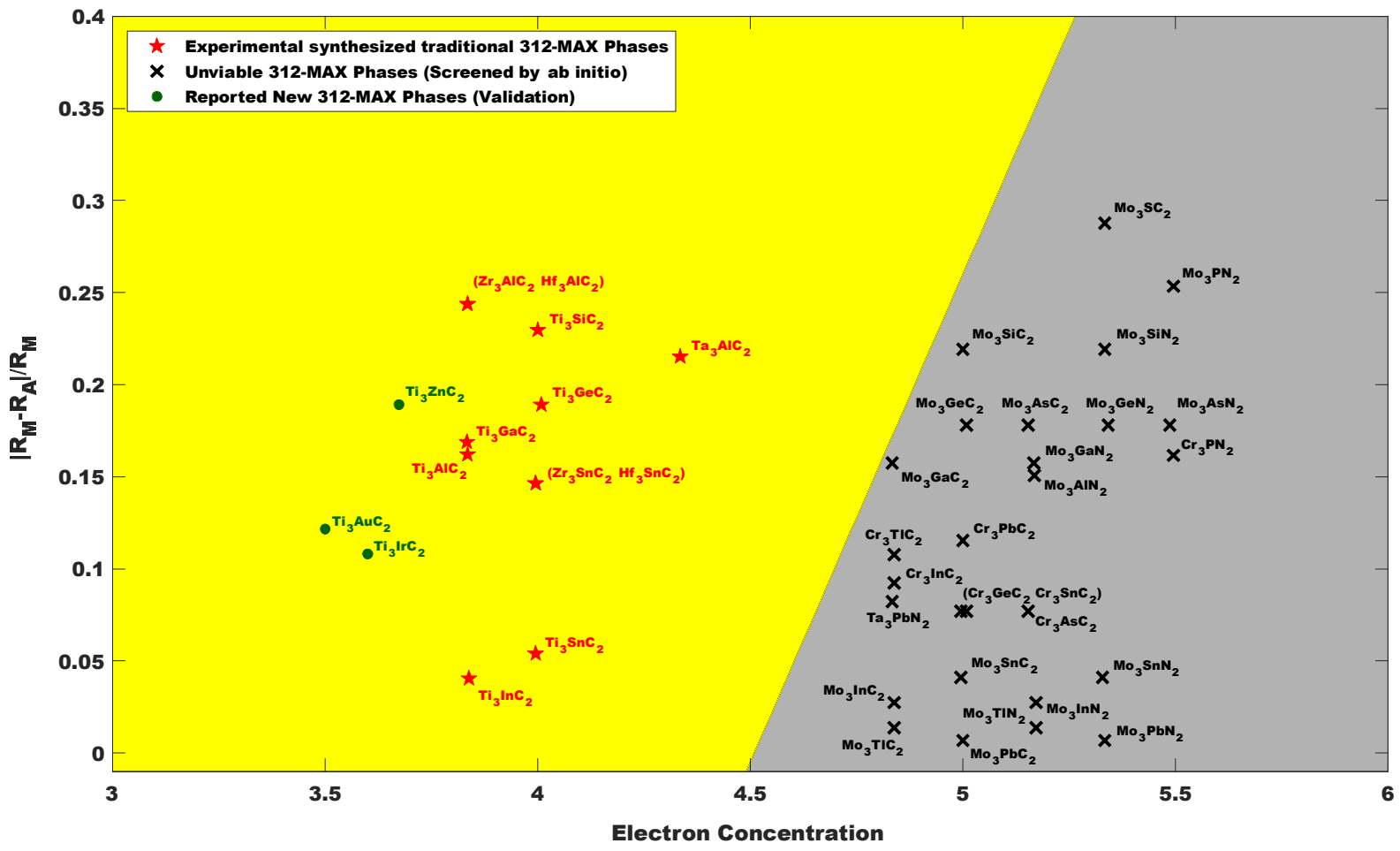

(b)

Figure 4 The constructed structure map further applied to (a) newly reported 211-MAX phases; (b) newly reported 312-MAX phases

### 3.4 *Employ the constructed structure map to explore potential MAX phases*

In order to guide future experimental development of MAX-phases, the developed structure map is employed to predict several potential MAX phases (shown in Figure 5).

We propose these marked potential 211-MAX phases could be discovered in future, several of which have been successfully synthesized [53, 56-59].

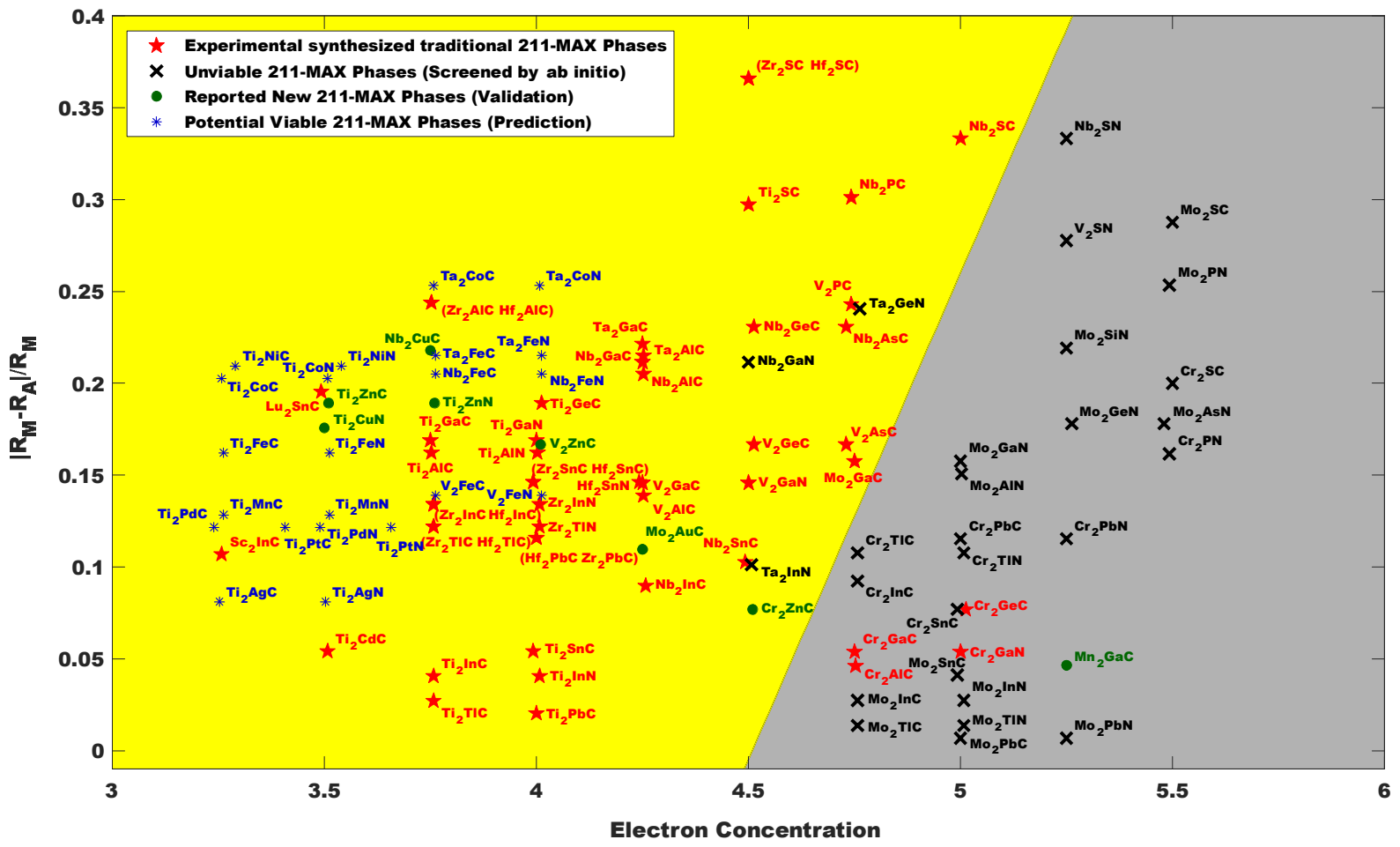

Figure 5 The prediction of future possible 211-MAX phases from the constructed structure map

## 4. Discussion

### 4.1 *The validity of electron concentration and size factor expression*

From the results shown for MAX phases formable/non-formable separation (shown in from Figure 2 to Figure 4), the *electron concentration factor* seems to play an important role in determining the MAX phases formability; which indicates the consistence of its choice with the standpoint raised by Barsoum [3], as well as the feasibility of its expression.

For the size factor, we choose the difference between atomic radii of M-site elements and A-site elements divided by radii of M-site elements as its expression. However, the atomic radii in compounding state are not precisely defined [40, 60]; and are affected by coordination number, oxidation state, or type of ligand [45]. For 312- and 413- MAX phases, there exist two different locations for M atoms; and allocating a single atomic diameter for each M-site element, independent of its respective environment, and neighbour atoms is too simplistic an approach. Further, in this work we only take the atomic size of M-site and A-site elements into consideration; without the ones of X-site elements. The formation of M-X block also plays role to the MAX phases formability.

### 4.2 *The contribution of other factors*

It has been mentioned that the structural stability is determined by four different groups. In this work, we have chosen *electron concentration factor* and *size factor* as controlling parameters to map associations of MAX phase structures to their compositions. Whereas the results show the feasibility of this choice, other factors can be introduced in to fully describe the MAX phases. For example, the electronegativity $\chi$ is a key feature in the description of the electronic interactions; and employing this parameter is proposed to enhance the description of covalent bonding characteristics within MAX phases.

### 4.3 *The MAX phases with elements of high electron concentration*

It is found that several of the synthesized Cr- and Mn- containing phases locate within the non-formable zone. This abnormality, connecting to high electron concentration of Cr and Mn, has been discussed in detail by Barsoum; where some of Cr-containing MAX phases (e.g. $Cr_2GeC$, $Cr_2AlC$) have been referred as anomalous or borderline-stable compounds [3].

However, it has to be pointed out that the size factor plays a role in confronting the destabilizing effect of electron concentration factor, when taking Mo- containing MAX phases as examples. The formability/non-formability of Mo- containing MAX phases are clearly separated; and the formation of $Mo_2GaC$ can be attributed to the size difference between Mo and Ga.

## 4. Conclusion

As one type of heterodesmic compounds, metallic, covalent and/or ionic interactions are present to a differing extent within MAX phases. To get a full description of crystal structures for it, regarding its chemical bonding and stability, the Schrödinger equations have to be solved. This strategy plays an important role in determining given compounds; still, the fact of computationally intensive characteristics limits the rate of potential new MAX phases investigation. In this work, we have shown the role that phenomenological structure map could take part in the search for future MAX phases via developing a classification scheme for MAX phases formability/non-formability separation. This compounding effect of *size factor* and *electron concentration factor* implicit within the constructed structure map could be utilized to guide the novel MAX phases design, either through M-site and A-site replacement, or via solid solution.

**Acknowledgement**

The authors acknowledge the financial support of National Natural Science Foundation of China (Grant No. 51872302, 21875271).

**Declaration of interest**

The authors declare that they have no known competing financial interests or personal relationships that could have appeared to influence the work reported in this paper.

## References

[1] M.W. Barsoum, The $M_{(N+1)}AX_{(N)}$ phases: A new class of solids; thermodynamically stable nanolaminates, Prog. Solid State Chem. 28 (2000), 201-281.

[2] Z.M. Sun, Progress in research and development on MAX phases: a family of layered ternary compounds, Int. Mater. Rev. 56 (2011), 143-166.

[3] M. W. Barsoum, MAX Phases: Properties of Machinable Ternary Carbides and Nitrides, Wiley-VCH: Weinheim **2013**.

[4] P. Eklund, M. Beckers, U. Jansson, H. Hogberg, and L. Hultman, The $M_{n+1}AX_n$ phases: materials science and thin-film processing, Thin Solid Films 518 (2010), 1851-1878.

[5] P. Eklund, J. Rosen, and P.O.A. Persson, Layered ternary $M_{n+1}AX_n$ phases and their 2D derivative MXene: an overview from a thin-film perspective, J. Phys. D: Appl. Phys. 50 (2017), 113001.

[6] M. Sokol, V. Natu, S. Kota, and M.W. Barsoum, On the Chemical Diversity of the MAX Phases, Trends Chem. 1 (2019), 210-223.

[7] M.F. Cover, O. Warschkow, M.M.M. Bilek, and D.R. McKenize, A comprehensive survey of $M_2AX$ phase elastic properties, J. Phy.: Condens. Matter 21 (2009).

[8] V.J. Keast, S. Harris, and D.K. Smith, Prediction of the stability of the $M_{n+1}AX_n$ phases from first principle*s*, Phys. Rev. B 80 (2009), 214113.

[9] M. Dahlqvist, B. Alling, and J. Rosen, Stability trends of MAX phases from first principles, Phys. Rev. B 81 (2010), 220102.

[10] S. Aryal, R. Sakidja, M.W. Barsoum, and W.-Y. Ching, A genomic approach to the stability, elastic, and electronic properties of the MAX phases, Phys. Status Solidi B 251 (2014), 1480-1497.

[11] D. Ohmer, I. Opahle, H.K. Singh, and H. Zhang, Stability predictions of magnetic $M_2AX$ compounds, J. Phy.: Condens. Matter 31 (2019), 405902.

[12] D. Ohmer, G. Qiang, I. Opahle, H.K. Singh, and H. Zhang, High-throughput design of 211-$M_2AX$ compounds, Phys. Rev. Mater. 3 (2019), 053803.

[13] D.G. Pettifor, Structure mapping, in Crystal Structures of Intermetallic Compounds, J. H. Westbrook and R. L. Fleischer, eds., John Wiley & Sons, New York, 2000, pp. 195-214.

[14] D.G. Pettifor, Structure maps revisited, J. Phy.: Condens. Matter 15 (2003), V13-V16.

[15] K. Rajan, Informatics for crystallography: designing structure maps, in Informatics for Materials Science and Engineering: Data-Driven Discovery for Accelerated Experimentation and Application, K. Rajan ed., Elsevier, Oxford, 2013, pp. 365-383.

[16] W. Hume-Rothery, The Engel-Brewer theories of metals and alloys, Prog. Mater. Sci. 13 (1967), 229-265.

[17] E. Mooser, and W.B. Pearson, On the crystal chemistry of normal valence compounds, Acta Crystallogr. 12 (1959), 1015-1022.

[18] J.C. Phillips, and Vanvecht.Ja, Spectroscopic Analysis of Cohesive Energies and Heats of Formation of Tetrahedrally Coordinated Semiconductors, Phys. Rev. B 2 (1970), 2147-2160.

[19] J.C. Phillips, and Vanvecht.Ja, Dielectric Classification of Crystal Structures, Ionization Potentials, and Band Structures, Phys. Rev. Lett. 22 (1969), 705-708.

[20] A. Zunger, Structural Stability of 495 Binary Compounds, Phys. Rev. Lett. 44 (1980), 582-586.

[21] P. Villars, A three-dimensional structural stability diagram for 1011 binary $AB_2$ intermetallic compounds: II, J. Less-Common Met. 99 (1984), 33-43.

[22] P. Villars, Three-dimensional structural stability diagrams for 648 binary $AB_3$ and 389 binary $A_3B_5$ intermetallic compounds: III, J. Less-Common Met. 102 (1984), 199-211.

[23] P. Villars, A three-dimensional structural stability diagram for 998 binary AB intermetallic compounds, J. Less-Common Met. 92 (1983), 215-238.

[24] D.G. Pettifor, The structures of binary compounds. I. Phenomenological structure maps, J. Phys. C: Solid State Phys. 19 (1986), 285-313.

[25] D.G. Pettifor, A chemical scale for crystal-structure maps, Solid State Commun. 51 (1984), 31-34.

[26] H. Zhang, N. Li, K. Li, and D. Xue, Structural stability and formability of $ABO_3$-type perovskite compounds, Acta Crystallogr., Sect. B: Struct. Sci. 63 (2007), 812-818.

[27] P.M. Clark, S. Lee, and D.C. Fredrickson, Transition metal $AB_3$ intermetallics: Structure maps based on quantum mechanical stability, J. Solid State Chem. 178 (2005), 1269-1283.

[28] J. Hauck, and K. Mika, Structure maps for crystal engineering, Cryst. Eng. 5 (2002), 105-121.

[29] A.S. Korotkov, and N.M. Alexandrov, Structure quantitative map in application for $AB_2X_4$ system, Comput. Mater. Sci. 35 (2006), 442-446.

[30] C.S. Kong, and K. Rajan, Rational design of binary halide scintillators via data mining, Nucl. Instrum. Methods Phys. Res., Sect. A 680 (2012), 145-154.

[31] C.S. Kong, P. Villars, S. Iwata, and K. Rajan, Mapping the 'materials gene' for binary intermetallic compounds-a visualization schema for crystallographic databases, Comput. Sci. Discovery 5 (2012), 015004.

[32] C.S. Kong, W. Luo, S. Arapan, P. Villars, S. Iwata, R. Ahuja, and K. Rajan, Information-theoretic approach for the discovery of design rules for crystal chemistry, J. Chem. Inf. Model. 52 (2012), 1812-1820.

[33] P. Villars, J. Daams, Y. Shikata, K. Rajan, and S. Iwata, A new approach to describe elemental-property parameters, Chem. Met. Alloys 1 (2008), 1-23.

[34] P. Villars, Factors governing crystal structures, in Crystal Structures of Intermetallic Compounds, J. H. Westbrook and R. L. Fleischer eds., John Wiley & Sons, New York, 2000, pp. 1-49.

[35] R. Ferro, A. Saccone, Intermetallic Chemistry, Elsevier, Amsterdan ; Oxford **2008**.

[36] W. Hume-Rothery, G.W. Mabbott, and K.M.C. Evans, The freezing points, melting points, and solid solubility limits of the alloys of silver and copper with the elements of the B sub-groups, Philos. Trans. R. Soc. London, Ser. A 233 (1934), 1-97.

[37] W. Hume-Rothery, Factors affecting the stability of metallic phases, in Phase Stability in Metals and Alloys, P. S. Rudman, J. Stringer and R. I. Jaffee eds., McGraw-Hill, New York, 1967, pp. 3-23.

[38] T.B. Massalski, Hume-Rothery rules re-visited, in Science of Alloys for the 21st Century: A Hume-Rothery Symposium Celebration, E. A. Turchi, R. D. Shull and A. Gonis eds., TMS (The Minerals, Metals & Materials Society), Warrendale, 2000, pp. 55-70.

[39] Y. Zhang, J.R.G. Evans, and S. Yang, The prediction of solid solubility of alloys: developments and applications of Hume-Rothery's Rules, J. Cryst. Phys. Chem. 1 (2010), 81-97.

[40] U. Mizutani, Hume-Rothery Rules for Structurally Complex Alloy Phases, CRC Press, Boca Raton, FL **2011**.

[41] G. Hug, M. Jaouen, and M.W. Barsoum, X-ray absorption spectroscopy, EELS, and full-potential augmented plane wave study of the electronic structure of $Ti_2AlC$, $Ti_2AlN$, $Nb_2AlC$, and $(Ti_{0.5}Nb_{0.5})_2AlC$, Phys. Rev. B 71 (2005), 024105.

[42] U. Mizutani, and H. Sato, The physics of the Hume-Rothery electron concentration rule, Crystals 7 (2017), 9-120.

[43] U. Mizutani, and H. Sato, Determination of electrons per atom ratio for transition metal compounds studied by FLAPW-Fourier calculations, Philos. Mag. 96 (2016), 3075-3096.

[44] D. Shriver, M. Weller, T. Overton, J. Rourke, F. Armstrong, Inorganic Chemistry, Oxford University Press, **2013**.

[45] M. Mantina, R. Valero, C.J. Cramer, and D.G. Truhlar, *Atomic radii of the elements*, in *CRC Handbook of Chemistry and Physics, 96th Edition (Internet Version 2016)*, W. M. Haynes ed., CRC Press/Taylor and Francis, Boca Raton, FL, 2016, pp. 9-49, 9-97.

[46] M. Ashton, R.G. Hennig, S.R. Broderick, K. Rajan, and S.B. Sinnott, Computational discovery of stable $M_2AX$ phases, Phys. Rev. B 94 (2016), 054116.

[47] M. Dahlqvist, and J. Rosen, Predictive theoretical screening of phase stability for chemical order and disorder in quaternary 312 and 413 MAX phases, Nanoscale 12 (2020), pp. 785-794.

[48] M. Li, J. Lu, K. Luo, Y. Li, K. Chang, K. Chen, J. Zhou, J. Rosen, L. Hultman, P. Eklund, P.O.A. Persson, S. Du, Z. Chai, Z. Huang, and Q. Huang, Element replacement approach by reaction with Lewis acidic molten salts to synthesize nanolaminated MAX phases and MXenes, J. Am. Chem. Soc. 141 (2019), 4730-4737.

[49] H. Ding, Y. Li, J. Lu, K. Luo, K. Chen, M. Li, P.O.Å. Persson, L. Hultman, P. Eklund, S. Du, Z. Huang, Z. Chai, H. Wang, P. Huang, and Q. Huang, Synthesis of MAX phases $Nb_2CuC$ and $Ti_2(Al_{0.1}Cu_{0.9})N$ by A-site replacement reaction in molten salts, Mater. Res. Lett. 7 (2019), 510-516.

[50] Y. Li, M. Li, J. Lu, B. Ma, Z. Wang, L.-Z. Cheong, K. Luo, X. Zha, K. Chen, P.O.A. Persson, L. Hultman, P. Eklund, C. Shen, Q. Wang, J. Xue, S. Du, Z. Huang, Z. Chai, and Q. Huang, Single-atom-thick active layers realized in nanolaminated $Ti_3(Al_xCu_{1-x})C_2$ and its artificial enzyme behavior, ACS Nano 13 (2019), 9198-9205.

[51] H. Fashandi, M. Dahlqvist, J. Lu, J. Palisaitis, S.I. Simak, I.A. Abrikosov, J. Rosen, L. Hultman, M. Andersson, A.L. Spetz, and P. Eklund, Synthesis of $Ti_3AuC_2$, $Ti_3Au_2C_2$ and $Ti_3IrC_2$ by noble metal substitution reaction in $Ti_3SiC_2$ for high-temperature-stable Ohmic contacts to SiC, Nat. Mater. 16 (2017), 814-818.

[52] S. Wang, J. Cheng, S. Zhu, J. Ma, Z. Qiao, J. Yang, and W. Liu, A novel route to prepare a $Ti_3SnC_2/Al_2O_3$ composite, Scr. Mater. 131 (2017), 80-83.

[53] Y. Li, J. Lu, M. Li, K. Chang, X. Zha, Y. Zhang, K. Chen, P.O.A. Persson, L. Hultman, P. Eklund, S. Du, Z. Chai, Z. Huang, and Q. Huang, Multielemental single-atom-thick A layers in nanolaminated $V_2(Sn,A)C$ (A=Fe, Co, Ni, Mn) for tailoring magnetic properties, PNAS 117 (2020), 820-825.

[54] C.-C. Lai, H. Fashandi, J. Lu, J. Palisaitis, P.O.A. Persson, L. Hultman, P. Eklund, and J. Rosen, Phase formation of nanolaminated $Mo_2AuC$ and $Mo_2(Au_{1-x}Ga_x)_2C$ by a substitutional reaction within Au-capped $Mo_2GaC$ and $Mo_2Ga_2C$ thin films, Nanoscale 9 (2017), 17681-17687.

[55] A.S. Ingason, A. Petruhins, M. Dahlqvist, F. Magnus, A. Mockute, B. Alling, L. Hultman, I.A. Abrikosov, P.O.A. Persson, and J. Rosen, A nanolaminated magnetic phase: $Mn_2GaC$, Mater. Res. Lett. 2 (2014), 89-93.

[56] Q. Huang, Y. Li, M. Li. One type of MAX phases, their processing strategy and applications. PCT Patent PCT/CN2018/117811

[57] Q. Huang, Y. Li, M. Li. A new ternary layered Max phase material with A-site as magnetic element. Chinese Patent ZL201810930369.0

[58] Q. Huang, Y. Li, M. Li. A kind of composite material of magnetic element composite ternary layered magnetic Max phase, its preparation method and application. Chinese Patent CN201910067712.8

[59] Y. Li, H. Ding, J. Lu, K. Chen, M. Li, P. O. Å. Persson, L. Hultman, P. Eklund, S. Du, Z. Chai, Z. Huang, Q. Huang, Sole magnetic element realized in nanolayered MAX (A=Fe, Co, Ni) phases via A-site replacement approach, to be submitted.

[60] Y.M. Zhang, S. Yang, and J.R.G. Evans, Revisiting Hume-Rothery's rules with artificial neural networks, Acta Mater. 56 (2008), 1094-1105.